\documentstyle[aps,multicol]{revtex}
\begin{document}\draft
\title{The Onsager Symmetry Relation and the Time\\ Inversion Invariance of 
the Entropy Production}
\author{Mario Liu}

\address{Institut f\"ur Theoretische Physik, Universit\"at Hannover,
      30167 Hannover, Germany, EC}
\date{\today}
\maketitle
\begin{abstract}
Starting from the entropy production being invariant under 
time reversal, one can (i) easily proof, and understand, many aspects of the 
linear Onsager relations and (ii) deduce the result that all quadratic 
Onsager coefficients for hydrodynamic fluxes vanish. 
\end{abstract}

\begin{multicols}{2}

Despite its usefulness, simplicity, and thorough experimental verification, 
the Onsager symmetry relation poses a number of vexing questions -- with 
respect to the ambiguity of its prescription, to its derivation and to its 
nonlinear generalization. Consider the temporal evolution of the entropy 
density, 
$\dot s- \nabla\cdot {\bf f}={\cal R}/T$, 
with $T$ denoting the temperature and ${\bf f}$ the entropy current; ${\cal 
R}$ is the source term, or entropy production, usually given as a sum of 
products: ${\cal R}=\sum_i\, b_iB_i$. Expanding the fluxes $b_i$ in the 
forces $B_i$ to linear order, $b_i=\sum_j\, L_{ij}B_j$,  the Onsager 
relation~\cite{GM,LL5} relates the kinetic coefficients pairwise, 
$L_{ij}={\bf tip}(B_i B_j)\, L_{ji}$, where ${\bf tip}(A)$ stands for the  
{\em time inversion parity} of the quantity $A$. The problems with this 
simple sounding prescription are: 

(1) On the most elementary level, given a product $b_1B_1$, it is not 
always clear which of the two is the force -- the time inversion property of 
which is so important for the Onsager relation.  This has been the main 
thrust of an eloquent and polemic criticism~\cite{Tr}. 

(2) A step further, the expression ${\cal R}=\sum_i\, b_iB_i$ is by no 
means unique, and the question naturally arises whether the Onsager relation 
still holds for ${\cal R}=\sum_i\,\tilde b_i\tilde B_i$, where $\tilde b_i$ 
and $\tilde B_i$ are linear combinations of $b_i$ and $B_i$. 

(3) In the original proof by Onsager, the fluxes are given by the temporal 
derivative of the state variables. This is not the case with hydrodynamic 
fluxes such as the entropy current ${\bf f}$ above; and it is not clear 
whether the symmetry relation holds here without further qualifications. On 
the other hand, we have no hint whatever of any invalidation experimentally;
and there have been successful efforts to bridge this gap for specific 
examples~\cite{GM,GC}. But a generally valid, clean-cut criterion for 
deciding when a given $b_i$ is a legitimate flux has proved elusive so far. 

(4) The validity of the Onsager relation is explicitly limited to linear 
terms, $b_i=\sum_j\, L_{ij}B_j$ with $L_{ij}$ constant. Nonlinear 
generalization appears difficult, and has focused on the question whether 
the symmetry relation such as it is holds nonlinearly, if $L_{ij}$ depends 
on $B_j$~\cite{nonlin}. 

In addition to these problems of the prescription, the Onsager relation 
has been the one result of the thermo- 
and hydrodynamic theory that could not be derived and justified within the 
framework of macroscopic physics. Moreover, the knowledge required for the 
excursion to microscopic dynamics and statistical mechanics is so advanced 
that many textbooks prefer to only state the results; others clad it in 
formidable notations and tuck it in the last few chapters of a thick book; 
the more convincing ones still fight the appearance of being at odds with 
the ultimate simplicity and universal validity of the Onsager relations. As 
a result, the knowledge about this relation has remained incommensurate with 
its fundamental significance and sheer usefulness; even the general 
awareness of it has suffered. 

So it must be a welcome discovery that there is 
in fact a simple and macroscopic derivation for the Onsager relation. 
Starting from the postulated invariance of the entropy production ${\cal R}$ 
under time inversion, only a few simple, algebraic steps are needed to 
conclude the proof, which includes the hydrodynamic fluxes and is easily 
extendable to nonlinear terms. It also eliminates the above ambiguities, 
showing that a sufficient condition for a bona fide force is an unambiguous 
{\bf tip}; and it specifies the most general linear transformation that 
leaves the Onsager relation undamaged. One such transformation actually 
reduces $L_{ij}$ to the unity matrix. 

Depending on sentiments, some will not want to call this a {\em proof}, as a new 
postulate is introduced. Nevertheless, since the invariance of ${\cal R}$ concisely 
summarizes all the above points, it is -- if true -- certainly a much better starting 
point when thinking about the Onsager relations, and when presenting them in 
books or lectures, as it should greatly facilitate the understanding and teaching of 
this chapter of irreversible thermodynamics. We should therefore at least accept it as 
a preliminary yet powerful hypothesis, while working to provide a proof, more 
rigorous than the considerations below -- or else, show where it may conceivably fail. 

In the following, the case for the high plausibility of ${\cal R}$'s invariance, 
${\bf tip}({\cal R})=+1$ is presented first. It is then employed, without further 
input, $\bullet$ to derive the Onsager relation, $\bullet$ to obtain the most general 
linear transformation, $\bullet$ to include hydrodynamic fluxes, and $\bullet$ to 
consider nonlinear ramifications, showing especially that all second order Onsager 
coefficients for hydrodynamic fluxes vanish. This is good news, as hydrodynamic 
theories --- such as the Navier-Stokes equations, the superfluid hydrodynamics of 
$^3$He~\cite{Kh}, the nematodynamics~\cite{dG}, and the hydrodynamic 
Maxwell equations~\cite{HyMax} --- are all an order more accurate 
then hitherto thought. 

Imagine particles in a box occupying only one half of the volume. Their velocity 
distribution is Maxwellian, with zero mean. Given these initial conditions, the 
particles will spread out quickly, and produce entropy on their way. If the time 
runs backward, $x\to x$, $v\to -v$ for each particle, and no macroscopic change in 
the initial conditions takes place. By pure phase space arguments we again expect 
the particles to spread. The rate of spreading, and with it the rate ${\cal R}$ of 
entropy production, must remain unchanged if the microscopic dynamics remains 
unchanged, if it is reversible. So, micro-reversibility leads straight to the invariance 
of ${\cal R}$. 

This simple scenario is of course a special case, but the conclusion should 
be generic, as the crucial element here, the reversibility of the 
 dynamics and initial conditions, holds widely: The first is 
believed to be rigorously true, the second presupposes local equilibrium, 
(or partial equilibrium where only a few slow macroscopic degrees of freedom 
are not in equilibrium). Off local equilibrium, say if we had more particles 
moving inwards than outwards initially, there will be a transient, ballistic 
period in which the cloud of particles will shrink, with the expansion 
resuming only afterwards. Reversing the time, the cloud will expand 
immediately. However, this temporal asymmetry disappears on a coarser time 
scale, on which the initial period is infinitesimally short. And only on 
this scale should the spreading be related to the (intrinsically 
coarse-grained) entropy production $\cal R$ -- which is zero in the 
transient range (no loss of information), and retains its invariance.

If there is an external magnetic field $H$, micro-reversibility is restored 
if the field is flipped with the time reversal operation. The same is true 
with  other macroscopic quantities that are odd under time inversion. 

Summarizing, we conclude that starting from any improbable, nonequilibrium 
state, the entropy increases for statistical reasons, in either directions 
of time. Given micro-reversibility, the rate of increase is the same. The 
first sentence implies ${\cal R}>0$, the second {\bf tip}(${\cal R})= +1$.  

To understand clearly the implication of these two expressions, consider 
$\dot s={\cal R}/T$ and its solution $s(t)$. Choose a point $t_0$ in time 
for a system that is off-equilibrium, ${\cal R}\not=0$. Note {\bf tip}($\dot 
s)=-1$ and {\bf tip}($T)= +1$.  If {\bf tip}(${\cal R})= -1$ were true, the 
solution $s(t)$ would be unchanged under the inversion of time, though the 
same curve is now traced backwards. With {\bf tip}(${\cal R})=+1$, inverting 
the time at $t_0$ reflects $s(t)$ at $t_0$. Retaining only the portion 
$t>t_0$ of $s(t)$, the two portions for either directions of time meet at 
$t_0$ and form a cusp there. With ${\cal R}>0$, the cusp points downwards. 
Accepting ${\cal R}>0$ yet allowing ${\cal R}=\sum_ib_iB_i= 
\sum_{ij}L_{ij}B_iB_j$ to contain terms for which {\bf tip}$(B_iB_j)= -1$, 
results in an asymmetric cusp, still pointing downwards but with the  left 
arm (depicting the time reversed solution) being flatter. 

Two final remarks. First, the proof below is such that it may be traced 
backwards, starting from the validity of the Onsager relation to demonstrate 
the invariance of $\cal R$. So, at least for the linear regime, close to 
global equilibrium, this invariance may be considered beyond dispute. At the 
same time, none of the considerations above makes any reference to how 
removed from global equilibrium the initial conditions may be. There is 
therefore no reason why ${\cal R}$ should not remain invariant including  
nonlinear Onsager coefficients. The half filled space is certainly far off 
equilibrium, not accountable by a linear expansion in density gradients. Yet 
this is our most straight forward example of an invariant ${\cal R}$. 

Second, when accounting for dilute systems with the kinetic theory,  higher 
order gradient terms are inherently related to higher order temporal 
derivatives. The latter account for inertial effects (or the dynamics of 
additional variables), and lead to ``temporal slips" such as the 
asymmetric spreading discussed above. In hydrodynamic theories of dense 
systems, these two expansions are quite independent. And an inclusion of 
higher order Onsager coefficients does not drive the system off local 
equilibrium. A comparison between the results of both theories must 
therefore be executed with circumspection~\cite{gold}. 

We proceed to consider the ramifications of an invariant ${\cal R}$, first 
in the context of non-hydrodynamic, global and spatially uniform variables. 
With $a_i$ a complete set of such variables,  and the entropy $S(a_i)$ 
maximal in equilibrium, we have (i) $A_i\equiv\partial S/\partial a_i=0$, 
and (ii) $\dot a_i=0$. Slightly off equilibrium, both $\dot a_i$ and $A_j$ 
are small, and we may expand $\dot a_i$ in $A_j$, leading to $\dot 
a_i=\sum_j\, L_{ij}A_j$. It is conventional to refer to $\dot a_i$ as the 
fluxes, and to $A_i$ as the forces. 

In the simplest case of one variable, $\dot S=\dot aA$, $\dot a=LA$, we 
find $R=LA^2=\dot a^2/L$, where $R\equiv\int{\rm d}^3r\,{\cal R}/T$. ($R$ is 
a function of independent variables, so it depends either on $A$ or on $\dot 
a$, but not on both simultaneously.) Clearly, irrespective of the ${\bf 
tip}(a)$, we have ${\bf tip}(R)=+1$. 

The same consideration for two variables is as far as we need to go: 
Starting with $\dot S=\dot a_1A_1+\dot a_2A_2$ and employing $\dot 
a_i=\sum_j\, L_{ij}A_j$, we obtain $R$ as a function of either $A_1$ and 
$A_2$, or $A_1$ and $\dot a_2$, 

\begin{eqnarray}
R&=&L_{11}A_{1}^2+L_{22}A_{2}^2+(L_{12}+L_{21})A_1A_2\label{R1}\\
&=&[DA_1^2+\dot a_2^2+(L_{12}-L_{21})A_1\dot a_2]/L_{22},\label{R2}
\end{eqnarray}
where $D\equiv L_{11}L_{22}-L_{12}L_{21}$. If $R$ is always to be even 
under time reversal, every single term of the above two sums has to 
be even, or vanish. Now, since $dS=A_ida_i$, ${\bf tip}(S)=+1$,  we have 
${\bf tip}(A_i)={\bf tip}(a_i)$, or ${\bf tip}(A_1\dot a_2)=-{\bf tip} 
(A_1A_2)$, where one of the two terms must be odd. And since the coefficient 
preceding the odd one has to vanish, we conclude: $L_{12}=-L_{21}$ if ${\bf 
tip}(A_1A_2) =-1$, and $L_{12}=L_{21}$ if ${\bf tip} (A_1\dot a_2)=-1$; or 
together $L_{12}={\bf tip}(A_1A_2)L_{21}$. As the subscripts 1 and 2 are 
arbitrary, we have already obtained the bulk of the Onsager relations. 
 
The presence of a static magnetic field $H$ is simple to include. 
If some kinetic coefficients depend on the field, say $\sim H$, the 
coefficients are themselves odd under time inversion. We shall denote these 
odd coefficients as $J_{ij}$, with ${\bf tip}(J_{ij})=-1$, and reserve 
$L_{ij}$ for the even ones. Substituting $L_{ij}$ by $J_{ij}$ in 
Eqs(\ref{R1},\ref{R2}), and reconsidering the {\bf tip}s of these terms 
yields two results: (i) All diagonal coefficients vanish, $J_{11}=J_{22}=0$. 
(ii) Off-diagonal coefficients are linked by: $J_{12}= -{\bf 
tip}(A_1A_2)J_{21}$. This concludes the proof of the Onsager relation. 

It is useful to adopt a more general view of the above algebra, and realize 
two points: (i) The facts that $\dot a_i$ is the time derivative of a 
thermodynamic variable, and $A_i$ the corresponding conjugate variable, were 
only used once, to deduce their respective {\bf tip}s. (ii) There is nothing 
essential to distinguish the forces from the fluxes. Both can be even or 
odd, both vanish in equilibrium, and both (being of the same number) 
represent a complete characterization of the nonequilibrium state. Hence we 
may expand $\dot a_i$ in $A_i$, or expand $A_i$ in $\dot a_i$, or in fact 
expand a mixture of both in the rest. With these two points in mind, it is 
obvious that we could have arrived at all the above conclusions by starting 
from $T\dot S=\sum_i\, b_iB_i$, knowing only ${\bf tip}(b_i)=-1$, ${\bf 
tip}(B_i)=+1$, and both vanish in equilibrium. We have especially

\begin{equation}
b_i=\textstyle{\sum_j}\,(L_{ij}+J_{ij})B_j,\ \mbox{with}\ L_{ij}=L_{ji},\ 
J_{ij}=-J_{ji}. \label{lOrel}\end{equation}

The generalization goes further: Any linear combinations of $b_i$ and $B_i$, 
yielding $\tilde b_i$ and $\tilde B_i$, will again lead to Eq(\ref{lOrel}) if the same 
conditions are met. Writing $\tilde b_i$ and $\tilde B_i$ as vectors, 
$\tilde{\bf b}= (\tilde{\bf L}+ \tilde{\bf J})\tilde{\bf B}$, we have 
$\tilde{\bf L}=\tilde{\bf L}^T$, $\tilde {\bf J}=-\tilde{\bf J}^T$, with 
$^T$ denoting the transposition operation, if $T\dot S=\tilde{\bf b}^T \tilde{\bf B}$, 
${\bf tip}(\tilde{\bf b}) = -1$, ${\bf tip}( \tilde{\bf B})=+1$ hold. All four 
transformations below satisfy these three conditions, 

\begin{eqnarray}
\tilde {\bf b}&=&{\bf \Theta}^T{\bf  b},\ \tilde {\bf B}={\bf \Theta}^{-1}{\bf B},
\ {\bf tip}{\bf (\Theta)}= +1,\\ 
\tilde {\bf B}&=& {\bf \Theta}^T {\bf b},\  \tilde{\bf b} ={\bf \Theta}^{-1}{\bf B}, 
\ {\bf tip}({\bf\Theta})=-1,\\
\tilde {\bf b}&=&{\bf b}+{\bf \Phi B},\ \tilde {\bf B}={\bf B},\\
\tilde {\bf B}&=&{\bf B}+{\bf \Phi b}, \ \tilde {\bf b}={\bf b},\\
{\bf\Phi}&=& -{\bf \Phi}^T,\ {\bf tip(\Phi)}= -1,\end{eqnarray} 
This completes a surprisingly brief discussion of transformations, 
nevertheless more general than those covered in the literature~\cite{GM,M}. 
Now, starting from Eq(\ref{lOrel}), we may take $\tilde{\bf b}={\bf b- JB}$, 
$\tilde{\bf B}={\bf B}$ to arrive at $\tilde{\bf b}= {\bf L}\tilde{\bf B}$. 
As $\bf L$ is symmetric, it can always be diagonalized. Dropping the tilde, 
a rescaling then reduces the dynamics to the strikingly simple form, 
${\bf b}={\bf B}$. 

Turning our attention now to hydrodynamic fluxes, we first consider heat 
transport in a stationary isotropic liquid with constant density, where the 
energy density only depends on the entropy density, ${\rm d}\varepsilon= 
T{\rm d}s$. (This corresponds to the one-variable case above.) 
Defining the energy flux as $T{\bf f}$, the energy conservation 
$\dot\varepsilon=\nabla\cdot (T{\bf f})$ is equivalent to 

\begin{equation}
\dot s-\nabla\cdot{\bf f}=({\bf f}\cdot\nabla T)/T.\label{dot s}
\end{equation}
Usually, taking ${\bf f}=\kappa\nabla T$, we accept the flux ${\bf f}$ and 
the source ${\cal R}=\kappa(\nabla T)^2$ as even under time reversal, 
opposite to $\dot s$, and hence as dissipative. This is certainly not wrong, 
but it shortcuts a subtle argument and obscures the perfect analogy to the 
nonhydrodynamic case above. The point is, the {\em intrinsic} time inversion 
parity for the flux is negative, ${\bf tip}({\bf f})\equiv {\bf tip}(\dot 
s)= -1$, and all three terms in Eq(\ref{dot s}) are odd, with the right side 
characterized by ${\bf tip}(\bf f)=-  {\bf tip}(\nabla T)$. Such as it 
stands, Eq(\ref{dot s}) is the exact counterpart to the starting equation of 
our previous consideration, $\dot S= \dot a\,A$, also a reversible equation, 
also with {\bf tip}($\dot a)= -{\bf tip}(A)$. Irreversibility is introduced in 
the next step, by equating these terms of different 
{\bf tip}s, in ${\bf f}=\kappa\nabla T$ as well as in $\dot a= L\,A$. 

The generalization to more variables is now clear: In all hydrodynamic 
theories~\cite{Kh,dG,HyMax}, we have   

\begin{equation}
\dot s-\nabla\cdot{\bf f}=(\sum_i\, b_iB_i)/T, 
\end{equation} 
where again ${\bf tip}(b_i)= -1$ and ${\bf tip}(B_i)= +1$, and 
both $b_i$ and $B_i$ vanish in equilibrium. This is because the force (such 
as $\nabla T$) has the {\bf tip} of the variable, while the 
flux (such as ${\bf f}$) shares the {\bf tip} with the time 
derivative of the variable. Hence the force-flux pair always has opposing {\bf tip}s,
and every single previous result is duplicated. 
 
We start the study of nonlinear Onsager coefficients with the one-variable 
case, $b=\sum_n (L_n+J_n)B^n$, where the subscript $_n$ denotes the 
$n^{th}$ order. Inserting the expression for $b$ in $R=bB$, we 
find $J_n=0$ for all $n$; switching then to $B=\sum_n\tilde L_nb^n$ we 
conclude $\tilde L_n=0$ for even $n$, or $B(b)=-B(-b)$. 

Before going on to more variables, we need to understand one crucial 
point. Expanding the trace of the momentum flux $\pi$ in the force 
$\nabla\cdot\mbox{\boldmath$v$}$, 
\begin{equation}
\pi/3=\textstyle{\sum_n}\,\eta_{2n+1} (\nabla\cdot\mbox{\boldmath$v$})^{2n+1}, 
\end{equation}
the viscosity coefficients 
$\eta_{2n+1}$ are functions of thermodynamic variables but do not depend on 
the force $\nabla\cdot\mbox{\boldmath$v$}$. Circumstances are 
different if we deal with relaxing variables, $\dot 
a=\sum_nL_{2n+1}A^{2n+1}$, $A=dS/da$. The coefficients $L_{2n+1}$ are again 
functions of thermodynamic variables, now including $a$, or equivalently $A$. 
The latter, 
however, is also a force. As there is no way to distinguish $A$-the-variable 
from $A$-the-force, this completely undermines the above result, allowing 
only odd orders in the force. Even if we admit only the linear 
term in the above sum, $\dot a=L_1(A)A$, we may still have terms of all orders 
in $A$. The electric current, $j=\sigma(E)E$, is a case in point: The 
existence of a rectifier alone shows that $j(E)$ cannot be an odd function 
of $E$~\cite{leit}. Therefore, this type of consideration will yield 
nonlinear order results only for those cases in which one can clearly 
distinguish the forces from the variables, as is true for all hydrodynamic 
theories. Conversely, counter examples from relaxative dynamics do not 
disprove the time reversal invariance of the entropy production. 

The insight that $\cal R$'s invariance has something to do with the Casimir 
part of the Onsager relation has probably crossed the mind of every 
practitioner of irreversible thermodynamics. It was Thomson who first 
proposed it. Meixner~\cite{M} dismisses it as accidental, because he overlooked 
the difference between the hydrodynamic and relaxing 
variables, and so was misled by the fact that some relaxing variables are 
known to possess quadratic terms.  

Going on to find second order cross coefficients, we choose the linearly 
diagonal representation, 

\begin{equation}
B_i=b_i+\textstyle{\sum_{ij}}(L_{ijk}+J_{ijk})b_jb_k, \label{quadr}
\end{equation}
with $L_{111}$, $L_{222}=0$, $L_{ijk}=L_{ikj}$; and the same for 
$J_{ijk}$. First, we neglect $J_{ijk}$ and confine the consideration to 
two variables. There are then four independent coefficients: 
$L_{112}=L_{121}$, $L_{122}$, $L_{212}=L_{221}$, and $L_{211}$. Inserting 
the equation of motion (\ref{quadr}) into the entropy production, 
$R=\sum_i\, b_iB_i$, we obtain 
$R=b_1^2+ L_1b_1^2b_2+b_2^2+ L_2b_1b_2^2$, 
where $L_1\equiv 2L_{112}+L_{211}$ and $L_2\equiv L_{122}+2L_{212}$ 
must be zero, as they precede odd terms. Substituting $B_1$ for $b_1$ in 
$R$, we find only one odd term, $B_1^2b_2$, the vanishing coefficient of 
which is $L_{211}-2L_{112}$, hence $L_{112}$, $L_{211}=0$. Substituting 
instead $B_2$ for $b_2$ exchanges the indices, leading to 
$L_{122}=2L_{212}$, and we conclude that all four $L$-coefficients 
vanish. By the same procedure, we find that the four $J$-coefficients 
vanish, too. Next, we include a third variable. Three independent even 
coefficients are in principle possible: $L_{123}=L_{132}$, 
$L_{213}=L_{231}$, $L_{321}=L_{312}$, in addition to the analogous odd ones. 
Again, various representations of $R$ leave no other choice than to have 
every single coefficient vanish. Finally, we note that in the consideration 
of quadratic Onsager coefficients, as there are at most three different 
variables in every term of $R$, we need not consider a fourth variable. This 
concludes the proof that no quadratic Onsager coefficient is compatible with 
the invariance of $R$. 

Cubic Onsager coefficients abound. Confining our attention to the  
two-variables case, taking $L_{ijkl}$ to be symmetric with respect to 
permutations of the last three indices, and again working in the linearly diagonal 
representation, the same procedure yields 6 independent coefficients: 
$3L_{1112}=L_{2111}$, $3L_{2221}=L_{1222}$, $L_{1111}$, 
$L_{2222}$, $L_{1122}$, and $L_{2112}$. 

The minimal conclusion one should draw from this paper is that the time 
inversion property of the entropy production $\cal R$ is worth paying much 
more attention to than is hitherto customary. This remains so even if one 
doubts the starting point of this paper, the invariance of $\cal R$, because 
much of the physics of dissipation is codified in  $\cal R$, in a highly 
condensed form. Bearing direct consequences on the symmetry relations 
between different kinetic coefficients, it is of great interests to those 
engaged in various hydrodynamic descriptions of complex fluids, and to those 
involved in kinetic theories. A specific example of how the kinetic theory, 
often carried out in oblivion to $\cal R$'s inversion property, would 
benefit from embracing it, is given in~\cite{karlin}.

\end{multicols}

\begin{thebibliography}{99}
\bibitem[*]{email} e-mail: liu@itp.uni-hannover.de

\bibitem{GM}S. R. de Groot and P. Masur, {\it Non-Equilibrium Thermodynamics},  
(Dover, New York 1984) \S IV3, VI4, VI5.

\bibitem{LL5}L.D. Landau  and  E.M. Lifshitz, {\it Statistical Physics},  
(Pergamon, Oxford, 1984), \S122;  D. Jou, J. Casas-Vazquez and G. Lebon, 
{\it Extended Irreversible Thermodynamics}, 2nd Ed. (Springer Verlag 1996) 

\bibitem{Tr}C. Truesdell, {\it Rational Dynamics} (Springer Berlin 1984) 
\S 7

\bibitem{GC}P. Goldstein and L.S. Garcia-Colin, Physica {\bf A242}, 467, 
(1997) 

 \bibitem{nonlin}J. Hurley and C. Garrod, Phys. Rev. Lett. {\bf 48}, 1575 
(1982); R. Rodriguez and L.S. Garcia-Colin, Phys. Rev. {\bf A36}, 4945 (1987); 
W. Brenig, {\it Statistical Theory of Heat}, (Springer Verlag, Berlin 1989) 
\S 7,15,16,17

\bibitem{Kh}N.D. Mermin and T.-L. Ho, Phys. Rev. Lett. {\bf 36}, 594 (1976);
C.-R. Hu and W.M. Saslow, Phys. Rev. Lett. {\bf 38}, 605 (1977); 
D. Lhuillier, J. Phys. Lett. {\bf 38}, 121 (1977); 
M. Liu and M.C. Cross, Phys. Rev. Lett. {\bf 41}, 250 (1978) and 
{\bf 43}, 296 (1979); M. Liu, Phys. Rev. Lett. {\bf 43}, 1740 (1979);
D. Vollhardt and P. W\"olfle,
{\it The Superfluid Phases of Helium 3}, Taylor and Francis, London (1990) 

\bibitem{dG} D. Forster, T.C. Lubenski, P.C. Martin, J. Swift and P.C. Pershan, 
Phys. Rev. Lett {\bf 26}, 1016 (1971); P.G. deGennes and J. Prost, 
{\it Physics of Liquid Crystals},  (Clarendon, Oxford, 1993); 
H. Pleiner and H.R: Brand, {\it Pattern Formation in Liquid Crystals}, 
Eds A. Buka and L. Kramer, (Springer, NY 1996) Ch. 2 

\bibitem{HyMax}M. Liu,  Phys. Rev. Lett. {\bf 70}, 3580 (1993), 
{\bf 74}, 1884, (1995), {\bf 74}, 4535 (1995), {\bf 77}, 1043 (1998); 

\bibitem{gold}By adding two variables, M.B. Romero and 
R.M. Valasco, Physica {\bf A222}, 161, 1995 achieved agreement 
between the Burnett coefficients and the Onsager relation.   

\bibitem{M} J. Meixner, Adv. Molec. Relax. Proc. {\bf 5}, 319, (1973)

\bibitem{leit}Taking $E\equiv{\bf n\cdot E}$ and $j\equiv{\bf n\cdot j}$ 
as the field and current along the easy direction ${\bf n}$, negative 
values are the same quantities in the ``hard'' direction. So $j(-E)=-j(E)$ 
implies the equivalence of both directions. 

\bibitem{karlin}The Bobylev instability of the Burnett and super-Burnett 
terms has been repaired recently by a cleverly executed infinite sum in a 
simplified model, rendering ${\cal R}$ positive for all wave lengths, see A. 
N. Gorban and I. V. Karlin Phys. Rev. Lett. {\bf 77}, 282, (1996); I. V. 
Karlin, G. Dukek and T. F. Nonnenmacher, Phys. Rev. E, {\bf 55}, 1573, 
(1997). Yet even this work yields an odd  term $\sim(\nabla_xv_x)^3$ in 
${\cal R}$. On the other hand, the starting equations of the simplified 
model do imply an ${\cal R}$ that is quadratic in the stress tensor and 
hence invariant. So the breaking of the invariance must be the result 
of algebraic manipulations, probably the truncation used to eliminate the 
stress tensor as a variable -- similar to the case in~\cite{gold} . 
\end{thebibliography}
\end{document}